# Application of deep learning to enhance the accuracy of intrusion detection in modern computer networks

**Jafar Majidpour, Hiwa Hasanzadeh**
[1]Dept. Computer Science. Raparin University Rania, Iraq

**ABSTRACT**

Application of deep learning to enhance the accuracy of intrusion detection in modern computer networks were studied in this paper. The identification of attacks in computer networks is divided in to two categories of intrusion detection and anomaly detection in terms of the information used in the learning phase. Intrusion detection uses both routine traffic and attack traffic. Abnormal detection methods attempt to model the normal behavior of the system, and any incident that violates this model is considered to be a suspicious behavior. For example, if the web server, which is usually passive, tries to There are many addresses that are likely to be infected with the worm. The abnormal diagnostic methods are Statistical models, Secure system approach, Review protocol, Check files, Create White list, Neural Networks, Genetic Algorithm, Vector Machines, decision tree. Our results have demonstrated that our approach offers high levels of accuracy, precision and recall together with reduced training time. In our future work, the first avenue of exploration for improvement will be to assess and extend the capability of our model to handle zero-day attacks.

*Corresponding Author:*

Jafar Majidpour,
Department Computer Science,
Raparin University Rania, Iraq.
Email: mailto:Jafar.majidpoor@uor.edu.krd

## 1. INTRODUCTION

Deep learning is a subcategory of machine learning and is based on a set of algorithms that attempt to model high-level abstract concepts in the data that process this process with The use of a deep graph, which has several layers of processing, consists of several layers of linear and nonlinear transformations. In other words, it's based on learning to display knowledge and features in the model layers. An educational sample (for example, a cat's image) can be modeled in a variety of forms, such as a mathematical vector filled with value per pixel, and more generally in the form of a set of smaller shapes (such as cat facial members). Some of these modeling techniques simplify the machine learning process (for example, cat image recognition) [1-3].

In the deep learning, there is hope to replace the extraction of these human-like features (such as cat members) with full automated observation and semi-monitoring techniques. The initial motivation for this learning structure has been inspired by the investigation of the neural structure in the human brain, those nerve cells allow perception by sending messages to each other. Depending on the various assumptions about how these neurons are connected, various models and structures have been proposed in this area, although these models do not naturally exist in the human brain, and the human brain has more complexity. hese models, such as Deep Neural Network, Complex Neural Network, Deep Intelligence Network, have made good progress in the areas of natural language processing, image processing. In fact, the deep learning vocabulary is the study of new methods for artificial neural networks [4-8].

In sum, deep learning is a sub-scan of machine learning that uses multiple layers of linear transformations to process sensory signals such as sound and image. The machine divides each complex concept into simpler concepts, and, with the continuation of this process, arrives at basic concepts that are able to make decisions for them, and thus there is no need for a complete human monitoring to specify the machine's information at any given moment. The subject matter of deep learning is important in providing information. Providing information to the car should be such that the car receives the key information it can decide upon by citing it in the shortest possible time. When designing deep learning algorithms, we must pay attention to the transformation factors that explain the information observed, these factors are not usually invisible, but factors that affect the visible handle or are the birth of human mental structures to simplify issues. For example, when processing speech, altering factors can be the speaker's dialect, age or gender. When processing a picture of a machine, the amount of glow in the sun is a metamorphic factor. One of the problems of artificial intelligence is the great effect of changing factors on received information. For example, many of the pixels received from a red car at night may be black. To solve these problems, we sometimes need to understand the information (about humans), and sometimes it's hard to find the right way to display information as much as it is [1-3, 9, 10].

The identification of attacks in computer networks is divided into two categories of intrusion detection and anomaly detection in terms of the information used in the learning phase. Intrusion detection uses both routine traffic and attack traffic. To accomplish this, various methods are used to enforce a series of illegal actions that compromise the integrity or access to a resource. Intrusions can be divided into internal and external categories. External influences are those that are inflicted by authorized or unauthorized persons inside the internal network from outside the network, and internal influences are made by authorized persons within the system and the internal network from within the network itself. Intruders generally use software deficiencies, password cracking, eavesdropping of network traffic and network design weaknesses, services, or network computers to penetrate computer systems and networks. An intrusion detection system can be set up. There are tools, methods, and documentation that help identify, identify, and report unregistered or unregistered activities under the network [5, 8, 10-14].

Intrusion detection is not a proper title for intrusion detection systems, because these systems really detect infiltration. They do not perceive the network activity as intrusion, which may not be essentially intrusive. In fact, cys Detection of penetration is a small part of the system's protective system and is not considered as an autonomous and independent system. Digital security tools can be considered equivalent to physical security tools. For example, the firewall 2 You can equate locked doors with an intrusion detection system equivalent to an alarm system and an intruder system as guardian dogs. Assume that you have a repository of confidential documents that you want to use with a fence around the premises, a system of announcements Protect your risk, locked doors, guard dogs and camera. Locked doors allow unauthorized people to enter It prevents the inside of the tank, but does not alert you in the event of an attacker's intrusion. The alarm system warns you, if the attacker intends to enter the tank, but does not prevent the penetration. Guards, It is an example of a series of measures that can prevent the intruder. As we have seen, the locked doors, the alarm system and the guard dogs, provided separate, but complementary roles in protecting the container of confidential documents. This was about the firewall, the intrusion detection and intrusion control system is also correct. These systems are different technologies that can work together. The location and layout of these tools can turn an unsecured network into a secure network. Intrusion detection systems are the task of identifying and detecting any unauthorized use of the system, exploitation and Or damage by both internal and external users. Intrusion detection systems are created as software and hardware systems, each with its own advantages and disadvantages. The speed and precision of the benefits of hardware systems, and the failure of their security failures by the attackers, is another feature of these systems. But the easy use of the software, the software compatibility and the differences in operating systems give more generality to software systems, and generally these systems are more appropriately chosen [6-9, 13-18].

Automatic Image Annotation is a technique or a tool to retrieve content-based and semantic concepts images. In technique, the image content is attached to a set of predefined switches. Content-Based Image Retrieval (CBIR) allows the users to retrieve the images efficiently. The image features are automatically extractable using image processing techniques. standardized color and texture called MPEG-7. These features include Color Layout Descriptor (CLD) and Scalable Color Descriptor (SCD) for colors and Edge Histogram Descriptor (EHD) for image texture [18-21].

Recently it has become essential to search for and retrieve high-resolution and efficient images easily due to swift development of digital images, many present annotation algorithms facing a big challenge which is the variance for represent the image where high level represent image semantic and low level illustrate the features, this issue is known as "semantic gab". This work has been used MPEG-7 standard to extract the features from the images, where the color feature was extracted by using Scalable Color Descriptor (SCD) and Color Layout Descriptor (CLD), whereas the texture feature was extracted by employing Edge Histogram Descriptor (EHD), the CLD produced high dimensionality feature vector therefor it is reduced by Principal Component Analysis (PCA). The features that have extracted by these three descriptors could be passing to the classifiers (Naïve Bayes and Decision Tree) for training [17, 22-25].

Figure 1 shows the intrusion detection system. As it can be seen in Figure 1, the server, Pc, switch, Firewall, IDS are the main components of the intrusion detection system.

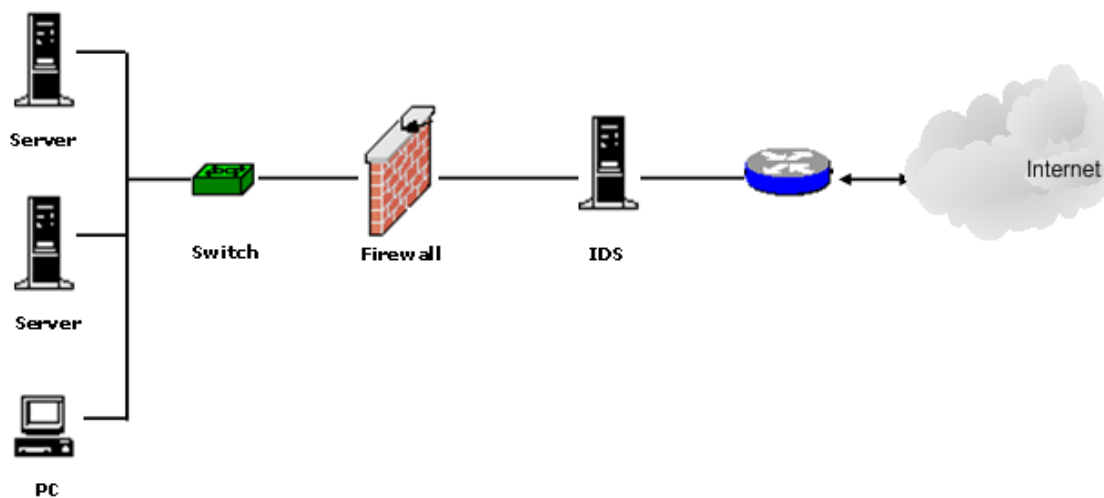

Figure 1. Intrusion detection system

## 2. RESEARCH METHOD

In today's world, computers and computer networks connected to the Internet play a major role in communications and information transfer. In the meanwhile, profitable individuals with access to important information of the special centers or other people's information with the intent of inflicting pressure or even pushing the system order violated the computer system Intruder and Cracker, Hacker Words that are nowadays more or less circulating in the computer system, affecting other systems and endangering their security. Therefore, the need to maintain information security and maintain the efficiency of computer networks that are connected with the outside world is quite tangible. Since technically creating computer systems (hardware and software) without the weaknesses and failures of security is practically impossible, Intrusion Detection is followed in the research of computer systems of particular importance. Intrusion detection systems have been used to help system security administrators to detect infiltration and attack. The purpose of an intrusion detection system is not to prevent attacks, and only detect and possibly detect attacks and detect security bugs in a system or computer networks and notify the system administrator. Generally, intrusion detection systems are used alongside the firewalls and complementary to them. Intrusion detection systems are essential for many organizations, from small offices to multinational corporations. Some of the benefits of this system are [6-9, 19, 26-30]:

a) More performance in intrusion detection, compared to manual systems
b) Source of complete knowledge of attacks
c) Ability to handle a large amount of information
d) Real-time alert ability that reduces damage
e) Give automatic answers, such as disconnecting the user, disabling the user's account, applying a set of automated commands, and so on
f) Increased deterrence
g) Ability to report

### 2.1. Intrusion Detection Methods

Detection methods used in intrusion detection systems are divided into two categories [27, 28]:
a) Abnormal behavior detection method
b) A method of detecting abuse or signature-based diagnosis.
　　Abnormal behavior detection:
　　To detect abnormal behavior, you must identify normal behaviors and find patterns and rules for them. Behaviors that follow these patterns are normal, and events that are more than usual statistically deviating from these patterns are recognized as abnormal behavior. Abnormal influences are very difficult to diagnose, since there is no steady pattern for monitoring. Usually, events that occur much more or less than two standard deviations from normal statistics are assumed to be abnormal. For example, if a user has to do it twenty times a day or two instead of one or two times a day, or a computer that was used at 2:00 after midnight while being placed. The computer has not been turned on after the office hours. Each of these can be considered as an abnormal behavior. The techniques and criteria used to detect abnormal behavior include [6, 30].:

A. Diagonal level detection

The number of log-outs to / from the system or the time of use of the system is a characteristic of the behavior of the system or user, which can be counted by its counting to the unusual behavior of the system, and it is due to a penetration. This is a totally static and explorative level of 8.

B. Commitment criteria

In a parametric type, the aggregated specification is considered based on a specific pattern and is compared in non-parametric mode based on the values obtained by the experiment. Famous IDSs that use statistical measurements to detect the abnormal behavior of penetration can be named 10NIDS.

C. Legislative criteria

Is similar to nonparametric statistical criteria, so that the observed data is defined by acceptable patterns in accordance with certain patterns. But it does not differ from the patterns that are specified by the law and are not indexed.

D. Other criteria

The first two criteria, namely threshold level detection and statistical criteria, are used in commercial IDSs. Unfortunately, detectors of abnormal influences and IDSs of this kind cause a lot of false alarms, which is because the patterns of behavior are very different from the users and the system. Instead, researchers claim that unlike signature-based diagnostic methods (which must necessarily coincide with previous patterns of attacks), abnormal behavior detection methods are capable of detecting a variety of new attacks. However, the creation of an intrusion detection system based on an abnormal behavior detection method is not always easy, and they do not have the accuracy of signature-based detection methods.

E. Signature detection

In this technique, pre-built penetration patterns (signatures) are kept in law. So that each pattern covers a different kind of specific influence and, if such a pattern is introduced in the system, it will be announced. In these methods, the detector usually has a database of signatures or attack patterns, and tries to find patterns similar to those stored in its database by checking the network traffic. These types of methods are only capable of detecting known infiltrations and can not detect them in the event of new network-level attacks, and the network administrator must always add the pattern of new attacks to the intrusion detection system. The advantages of this method are precision in the detection of infiltration, which their model has given to the system [5-15, 26, 27].

**2.2. Introduction of Different Types of Intrusion Detection in Abnormal Detection**

Abnormal detection methods attempt to model the normal behavior of the system, and any incident that violates this model is considered to be a suspicious behavior. For example, if the web server, which is usually passive, tries to There are many addresses that are likely to be infected with the worm. The abnormal diagnostic methods discussed here are:

a) Statistical models
b) Secure system approach
c) Review protocol
d) Check files
e) Create whitelist
f) Neural networks
g) Genetic algorithm
h) Vector vector machines
i) Decision tree.

**2.3. Statistical Models**

Statistical models are among the first methods used to detect infiltration. In this way, it is assumed that the attacking behavior is appropriate to the normal user. Similarly, for other topics, such as groups and programs, it is used. Some of the statistical characteristics of events that are used to diagnose are:

a) Threshold size: This method attributes value or originality to trigger events or how many times they occur over a period of time. Some common examples of user logins and user deactivation after Number of failures in the entry

b) Medium and standard deviation: By comparing events with a median profile and standard deviation, the confidence interval for anomalies is estimated. Profiles are based on historical data or pre-configured values.

c) Multivariate models: Calculation of the correlation between different incident scales, according to the expectations of the profile.

d) Markov Process Model: This model considers types of attacks as state variables in a state / pass state matrix. In this system, an abnormal occurrence is considered, if the probability of occurrence of that occurrence for the previous state with its dependent value, Very low.

e) Classification analysis: This non-parametric method operates in a vector based on the flow of events, which are grouped using classifying algorithms in different classes of behaviors. The subdivisions include the same or the patterns are user-friendly, so that the normal behavior of the abnormal is distinct.

**2.4. Secure System Approach**

In essence, applications implement a model of normal program behavior through code execution. In a secure system approach, applications are modeled according to a sequence of system calls for different conditions (including normal behavior, state Error and attempts to exploit) Comparing this model with the effects of observed events provides the possibility to categorize normal and abnormal behavior.
For example, abnormal behavior of an executable call on a web server can represent a stack overflow attack. This method is capable of detecting many common attacks, but it is not possible to detect attacks that are based on match conditions, policy violations, or tampering.

**2.5. Protocol Review**

Many of the attack methods are based on unusual or abnormal use of protocol fields. Protocols are checking the accuracy of the fields and behavior of the protocol in accordance with standards and expectations. Data that violates this range is considered as suspicious data. This approach identifies many common attacks, but the underlying problem is the negligent observance of standards in the implementation of many protocols. Also, the use of This way, for proprietary or unfamiliar protocols, can create false alarms.

**2.6. Neural Network**

Neural networks can be called with great negligence electronic models of the neural structure of the human brain and the basis of neural analytical models is based on the simulation of the activities of a neuron (neuron), the brain as the information processing system of a large number of neurons (neurons). The simplest structural unit of the nervous system is composed. The neural network is a software program that can act as a human brain, so that it can be more efficient over time and more interacting with the environment; in addition to making calculations, it is possible to make a logical conclusion; (Generalizability). Neural networks are widely used as an effective method for classifying patterns matching, but high volume computing and long learning cycles are far behind in many applications. In detecting maladaptive infiltration (unknown attacks) An artificial neural network is used, artificial neural network to identify unknown attacks and to avoid malicious intrusions has more advantages than other methods (based on statistics and rule-based) [11, 12].

The intrusion detection system based on neural networks for a specific computer system includes the following three steps:
a) Training Datasheets: Getting logs for each user in multi-day periods for each user through a vector shows what a user is doing.
b) Training: Neural network to identify the user based on the commands in the vector.
c) Performance: The network identifies the user for each new command, which means that if the user executes a new command that does not exist in his vector, the system will be able to identify that user.

**2.7. Genetic Algorithm**

Here are two phases. The first phase is training, which teaches the basic information required by the system. The second phase recognizes that the system detects infiltration based on first-phase training of infiltrations. In intrusion detection systems that use genetic algorithms for training, we put a series of basic rules into a database, and by applying the genetic algorithm, new rules are created and added to the preceding rules. In the figure below, the structure of a simple genetic algorithm is shown.

Genetic algorithms use genetic evolution as a problem solving model. Solutions are coded according to a pattern that evaluates the function of each candidate solution, most of which is called fitness. The evolution of a completely randomized collection of early society begins and is repeated in subsequent generations. This process is repeated until we reach the last step. The termination conditions of the genetic algorithm can be as follows:
a) We reach a fixed number of generations.
b) Allotted funds
c) The highest degree of fit of children or other results
d) Do not get better
e) Manual inspection

f) Combinations of items listed above [5-15].

As it dominstrated in Figure 2, "generate initial population" is the first step of the Generic algorithm. After that, as a second step the objective function is evaluatedthe the system should answer this question: "Are optimizataion critreria met?" If the answer is "Yes" the best individuals are saved as a result.

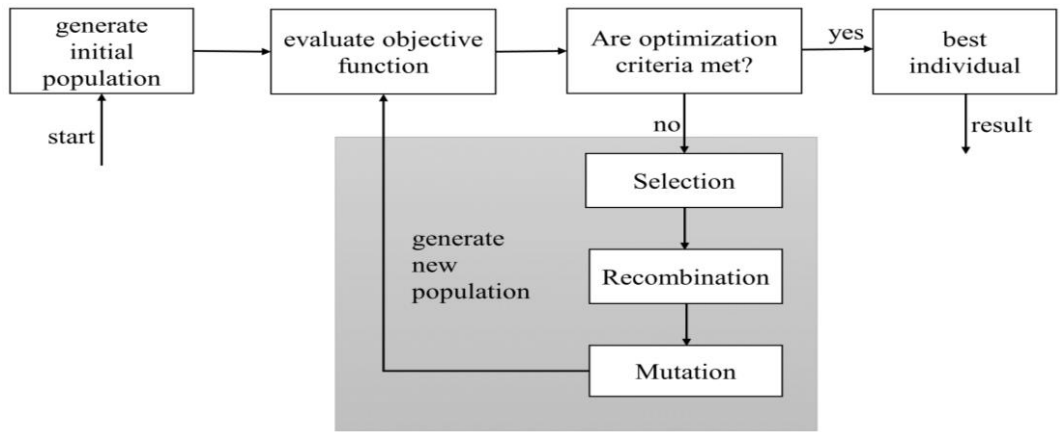

Figure 2. Generic algorithm

## 3. RESULTS
### 3.1. Introducing a Variety of Intrusion Detection Architectures

This system detects and detects unauthorized activity on the host computer. A host-based intrusion detection system can detect attacks and threats on critical systems (including access to files, trojan horses, etc.) that can not be detected by network-based intrusion detection systems. HIDS protects only the hosts on which they are based, and their network interface card (NIC) works by default in a steady state. The mode with the rule can be useful in some cases. Because all the network interface cards do not have inaccurate 19-bit mode. Due to their host location, which must be monitored, HIDS are aware of all kinds of additional local information with security implications (including system calls, system file changes, and system connections). When combined with network communications, this issue provides good data to search for possible events. The other advantage of HIDS is the ability to organize very good decisions for each unique host. For example, there is no need for a host that does not run Domain Name Servers (DNS), multiple rules that are designed to detect malware from DNS. As a result, reducing the number of relevant rules will increase efficiency and reduce the overhead of the processor for each host. HIDS also provides specific information on where, where and by whom, where and by whom, influence is taking place. This operation is very useful because there is no shortage or deletion. In host-based IDS, the possibility of false alarms is very small, since the information goes directly to the users of the applications. These IDSs have less traffic than NIDS, with more emphasis on multiple multicore sensors and central management stations. One of the disadvantages of HIDS is the low compatibility between the operating system and the resulting multiple software. Most host-based IDSs are written only for an operating system. The other is that HIDS does not identify some of the attacks that occur in the bottom layers of the network.

Figure 3 shows the host-based intrusion detection system which the firewall, HIDS, web servers and mail server are the main components of this system.

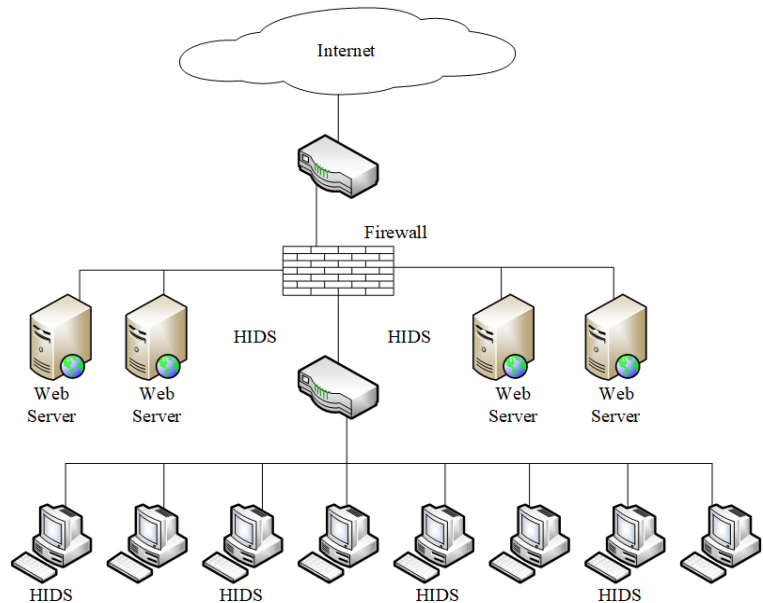

Figure 3. Host-based intrusion detection system

The NIDS name is derived from the fact that it monitors the entire network from a local perspective. Identifying and detecting unauthorized influences Before reaching critical systems, they are responsible for network-based intrusion detection system. NIDSs are often composed of two parts of the sensor (sensor) and agent. These two parts are often installed behind the firewall and the rest of the access points to detect any unauthorized activity. Network agents can replace the network infrastructure to search for network traffic. Installing agents and observers has the advantage of eliminating any kind of attack initially. Meanwhile, the sequence of reviews by one or more hosts can be useful to look for the symptoms of attacks.

Generally, the computer's network card works in solid state. In this case, only the packets that are sent to the physical network (MAC) address of the network card (or all-packets) are sent to the protocol stack for analysis. The NIDS should work in an irregular state to address the network traffic sent to the address MAC itself is not monitored. In idle mode, the NIDS can eavesdrop on all communications in network segments. The operation of the NIDS in an idle mode is essential for network protection. Although, from the point of view of confidentiality and eavesdropping, surveillance of network communications is a responsibility that needs to be carefully addressed. It shows a network of four NIDS utilities. Units are placed on strategic segments and can monitor network traffic for all segment segments. This configuration provides a standard for network security topology, so that separate subnetworks covered by public servers are protected by NIDS. NIDS can be programmed to prevent interruptions during the work. So that the presence of any shipments that NIDS detects is recorded inside the event file and notifies the network administrator without the attacker's knowledge. Network-based intrusion detection systems require a password for applications, salaries It is not related to the network operating system or system connections when running the software. Also, because these systems operate at the network layer level, they are not affiliated with the operating system. Meanwhile, there is no overhead on servers and workstations, because for these intrusion detection systems there is no need to install additional tools.

These systems consist of several NIDS or HIDS or a combination of these two types together with a central management station. Each IDS on the network sends its reports to the central management station. The central station is responsible for reviewing reports and notifying the system security officer. This central station is also responsible for updating the rules of the detection of each IDS on the network. The subtle form of the distributed penetration detection system is displayed. NIDS is responsible for protecting public servers and NIDS 3,4 to protect the internal network. Information is stored at the central management station. The network between the NIDSs with the central management system can be private or that the existing infrastructure for data transmission is used. When an existing network is used to send management data, additional security is achieved by cryptography or virtual private VPN technology [5-15].

### 3.2. Exploring the Use of Deep Learning to Detect Network Intrusion

Network monitoring has been used extensively for different detection. However, recent advances have created many new obstacles for NIDSs. Some of the most pertinent issues include:

a) Volume - The volume of data both stored and passing through networks continues to increase. It is forecast that by 2020, the amount of data in existence will top 44 ZB. As such, the traffic capacity of modern networks has drastically increased to facilitate the volume of traffic observed. Many modern backbone links are nowoperating at wirespeeds of 100 Gbps or more. To contextualise this, a 100 Gbps link is capable of handling 148,809,524 packets per second. Hence, to operate at wirespeed, a NIDS would need to be capable of completing the analysis of a packet within 6.72 ns. Providing NIDS at such a speed is difficult and ensuring satisfactory levels of accuracy, effectiveness and efficiency also presents a significant challenge.

b) Accuracy - To maintain the aforementioned levels of accuracy, existing techniques cannot be relied upon. Therefore, greater levels of granularity, depth and contextual understanding are required to provide a more holistic and accurate view. Unfortunately, this comes with various financial, computational and time costs.

c) Diversity - Recent years have seen an increase in the number of new or customised protocols being utilised in modern networks. This can be partially attributed to the number of devices with network and/or Internet connectivity. As a result, it is becoming increasingly difficult to differentiate between normal and abnormal traffic and/or behaviours.

d) Dynamics - Given the diversity and flexibility of modern networks, the behaviour is dynamic and difficult to predict. In turn, this leads to difficulty in establishing a reliable behavioural norm. It also raises concerns as to the lifespan of learning models.

e) Low-frequency attacks - These types of attacks have often thwarted previous anomaly detection techniques, including artificial intelligence approaches. The problem stems from imbalances in the training dataset, meaning that NIDS offer weaker detection precision when faced with these types of low frequency attacks.

f) Adaptability -Modern networks have adopted many new technologies to reduce their reliance on static technologies and management styles. Therefore, there is more widespread usage of dynamic technologies such as containerisation, virtualisation and Software Defined Networks. NIDSs will need to be able to adapt to the usage of such technologies and the side effects they bring about.

An example of an encoder auto and an example of an encrypted auto encoder are shown below.
As it can be seen in Figure 4, the hidden layer in the auto-encoder is a layer in between input layers and output layers, where system take in a set of weighted inputs and produce an output through an activation function. Figure 5 shows an example of a stacked auto-encoder which the hidden layers are trained by an unsupervised algorithm and then fine-tuned by a supervised method.

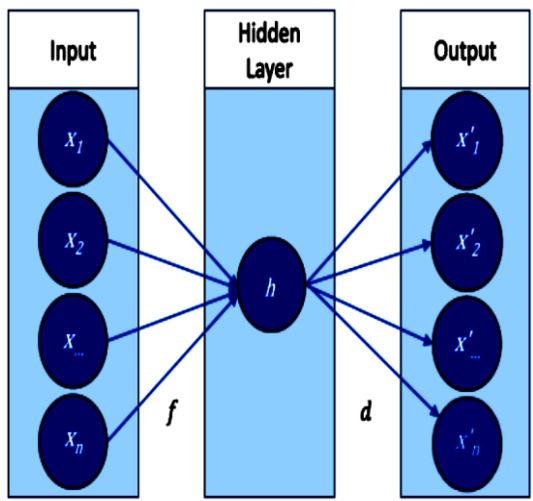
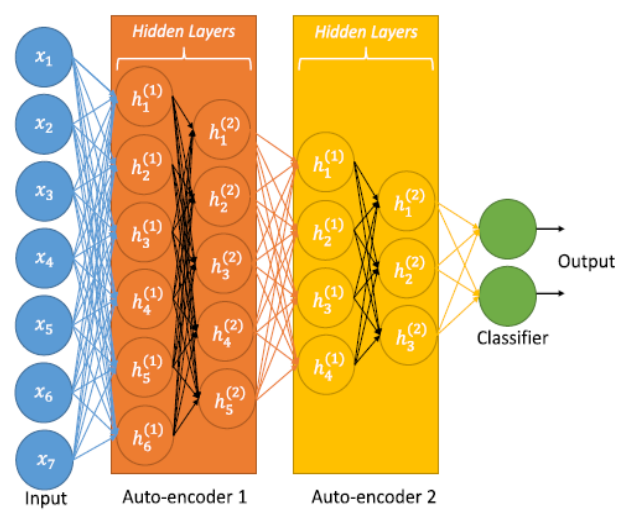

Figure 4. An example of a single auto-encoder    Figure 5. An example of a stacked auto-encoder

**3.3. Non-Symmetric Deep Auto-Encoder**

Decreasing the reliance on human operators is a crucial requirement for future-proofing NIDSs. Hence, our aim is to devise a technique capable of providing reliable unsupervised feature learning, which can improve upon the performance and accuracy of existing techniques. NDAE can be used as a hierarchical unsupervised feature extractor that scales well to accommodate high-dimensional inputs. It learns non-trivial features using a similar training strategy to that of a typical auto-encoder. An illustrated example of this is presented in below.

NDAE can be used as a hierarchical unsupervised feature extractor that scales well to accommodatehigh-dimensional inputs. It learns non-trivial features using a similar training strategy to that of a typical auto-encoder. An illustrated example of this is presented in Figure 6.

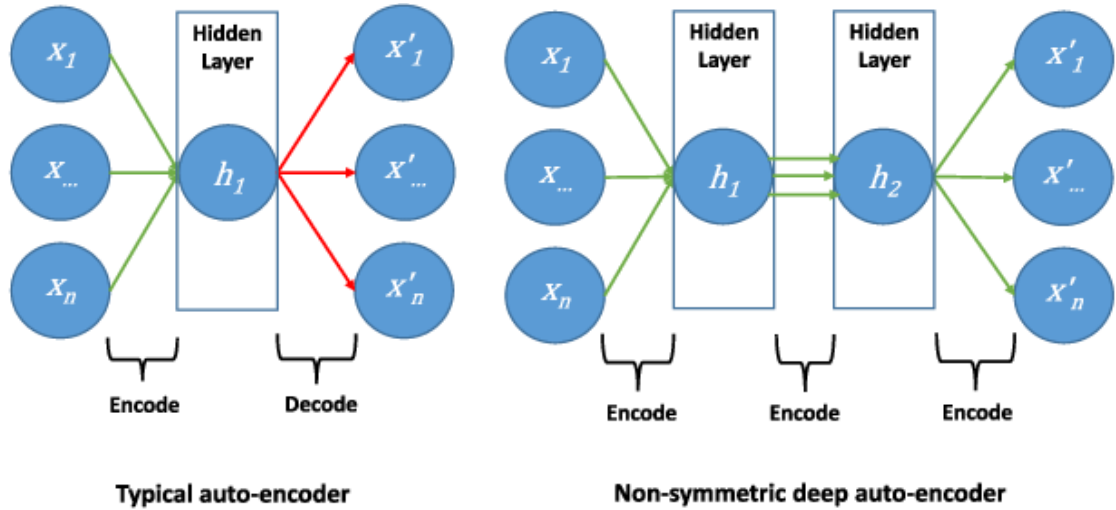

Figure 6. Comparison of a typical auto-encoder and a NDAE

**3.4. Stacked Non-Symmetric Deep Auto-Encoders**

In this subsection, we detail the novel deep learning classification model we have created to address the problems identified with current NIDSs. Fundamentally, our model is based upon using our NDAE technique for deep learning. This is achieved by stacking our NDAEs to create a deep learning hierarchy. Stacking theNDAEs offers a layer-wise unsupervised representation learning algorithm, which will allow our model to learn the complex relationships between different features. It also has feature extraction capabilities, so it is able to refine the model by prioritising the most descriptive features. Due to the data that we envisage this model using, we have designed the model to handle large and complex datasets. Despite the features present in the KDD Cup '99 and NSL-KDD datasets being comparatively small, we maintain that it provides a benchmark indication as to the model's capability. However, the classification power of stacked auto-encoders with a typical soft-max layer is relatively weak compared to other discriminative models including RF, KNN and SVM.

In deep learning research, the exact structure of a model dictates its success. Currently, researchers are unable to explain what makes a successful deep learning structure. The exact structure of our model has resulted from experimented with numerous structural compositions to achieve the best results. The final structure of our proposed model is shown in Figure 7.

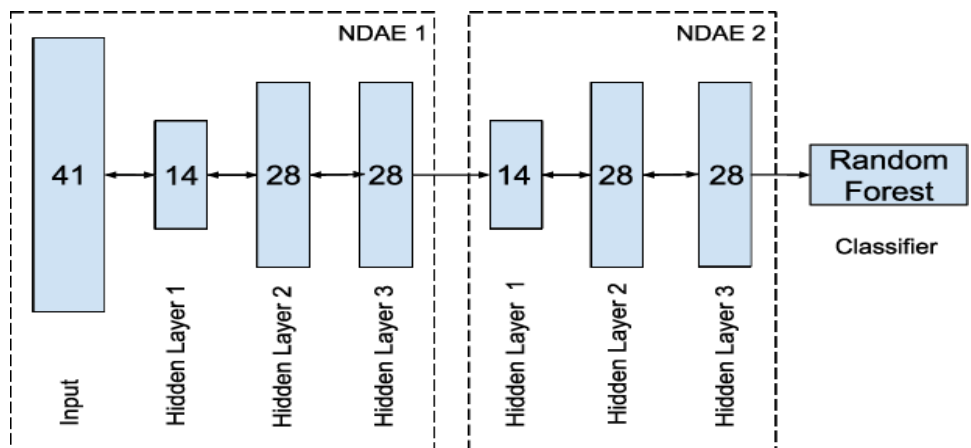

Figure 7. Stacked NDAE classification model

RF is basically an ensemble learning method, the principle of which is to group 'weak learners' to form a 'strong learner'. In this instance, numerous individual decision trees (the weak learners) are combined to form a forest. RF can be considered as the bagging (records are selected at random with replacement from the original data) of these un-pruned decision trees, with a random selection of features at each split. It boasts advantages such as low levels of bias, robustness to outliers and overfitting correction, all of which would be useful in a NIDS scenario.

a) True Positive (TP) - Attack data that is correctly classified as an attack.
b) False Positive (FP) - Normal data that is incorrectly classified as an attack.
c) True Negative (TN) - Normal data that is correctly classified as normal.
d) False Negative (FN) - Attack data that is incorrectly classified as normal.

We will be using the following measures to evaluate the performance:

$$Accuracy = \frac{TP+TN}{TP+TN+FP+FN} \quad (1)$$

The accuracy measures the proportion of the total number of correct classifications.

$$Precision = \frac{TP}{TP+FP} \quad (2)$$

The precision measures the number of correct classifications penalised by the number of incorrect classifications.

$$Recall = \frac{TP}{TP+FN} \quad (3)$$

The recall measures the number of correct classifications penalised by the number of missed entries.

$$False\ Alarm = \frac{FP}{FP+TN} \quad (4)$$

The false alarm measures the proportion of benign events incorrectly classified as malicious.

$$F-score = 2.\frac{Precision.Recall}{Precision+Recall} \quad (5)$$

The F-score measures the harmonic mean of precision and recall, which serves as a derived effectiveness measurement. This paper utilises the KDD Cup '99 and NSL-KDD benchmark datasets. Both of which have been used extensively in IDS research involving traffic with both normal and abnormal connections. The results shown Table 1. It should be noted that the values of the following table are normalized on the basis of statistical rules and the evaluation of errors, and these values are considered as final values.

Table 1. KDD CUP '99 Performance

| Attack class | No. training | No. attacks | Accuracy | | percision | | Recall | | F-score | | False alarm | |
|---|---|---|---|---|---|---|---|---|---|---|---|---|
| | | | DBN | S-DANE | DBN | S-DANE | DBN | S-DANE | DBN | S-DANE | DBN | S-DANE |
| Normal | 97260 | 59580 | 98.55 | 98.45 | 93.32 | 99.5 | 98.56 | 98.35 | 95.53 | 98.45 | 4.57 | 8.35 |
| Dos | 391440 | 223190 | 98.32 | 98.15 | 97.56 | 98.96 | 98.76 | 98.25 | 98.72 | 98.46 | 1.15 | 0.035 |
| Probe | 4100 | 2340 | 13.81 | 97.56 | 85.46 | 99.32 | 13.85 | 96.45 | 23.48 | 98.54 | 12.65 | 10.56 |
| R2l | 1115 | 5980 | 88.15 | 8.53 | 99.94 | 99.24 | 88.45 | 8.52 | 93.56 | 16.98 | 0.00 | 0.62 |
| U2r | 48 | 32 | 6.93 | 0.00 | 36.67 | 0.00 | 6.45 | 0.00 | 11.98 | 0.00 | 59.97 | 99.99 |
| Total | 493056 | 291255 | 96.85 | 96.57 | 96.57 | 99.15 | 96.68 | 96.56 | 96.34 | 97.89 | 1.98 | 1.98 |

## 4. CONCLUSION

Application of deep learning to enhance the accuracy of intrusion detection in modern computer networks were studied in this paper. Abnormal detection methods attempt to model the normal behavior of the system, and any incident that violates this model is considered to be a suspicious behavior. For example, if the web server, which is usually passive, tries to There are many addresses that are likely to be infected with the worm. The abnormal diagnostic methods are: Statistical models, Secure system approach, Review protocol, check files, Create Whitelist, Neural Networks, Genetic Algorithm, Vector Vector Machines, decision tree.

Our results have demonstrated that our approach offers high levels of accuracy, precision and recall together with reduced training time. In our future work, the first avenue of exploration for improvement will be to assess and extend the capability of our model to handle zero-day attacks.

## REFERENCES

[1]   H. A. Song and S.Y. Lee, "Hierarchical Representation Using NMF," *Neural Information Processing*, Springer Berlin Heidelberg, 2013.
[2]   B. A. Olshausen, "Emergence of simple-cell receptive field properties by learning a sparse code for natural images," *Nature 381.6583*, pp. 607-609, 1996.
[3]   R. Collobert, "Deep Learning for Efficient Discriminative Parsing," *videolectures.net. Ca*, vol. 7, pp. 45, 2011.
[4]   L. Gomes, "Machine-Learning Maestro Michael Jordan on the Delusions of Big Data and Other Huge Engineering Efforts," *IEEE Spectrum*, Oct 2014.
[5]   P. Innella and O. McMillan, "An Introduction to Intrusion Detection Systems," 2001.
[6]   P. Ning and S. Jajodia, "Intrusion Detection Techniques," *Intrusion Detection Techniques*, 2004.
[7]   B. Caswell, J. Beale and A. Baker, "Snort IDS and IPS Toolkit," *SyngressPublishing*, 2007.
[8]   M. Analoui, A. Mirzaei and P. Kabiri, "*Intrusion detection using multivariate analysis of variance al-gorithm*," In Third International Conference on Systems, Signals & Devices SSD05, vol. 3. 2005.
[9]   V. Kumar and O. P. Sangwan, "Signature based intrusion detection system using SNORT," *International Journal of Computer Applications & Information Technology*, vol. 1, no. 3, pp. 35-41, Nov 2012.
[10]  P. Kabiri, and A. Ghorbani. "Research on Intrusion Detection and Response: A Survey," *IJ Network Security,* vol. 1, no. 2, pp. 84-102, 2005.
[11]  V. Nourani, *et al*., "Emotional artificial neural networks (EANNs) for multi-step ahead prediction of monthly precipitation; case study: northern Cyprus," *Theoretical and Applied Climatology*, pp. 1-16, 2019.
[12]  V. Nourani, *et al*., "ANN-based statistical downscaling of climatic parameters using decision tree predictor screening method," *Theoretical and Applied Climatology*, vol. 137, pp. 1729-1746, 2019.
[13]  Lixin Wang, "Artificial Neural Network for Anomaly Intrusion Detection," 2003.
[14]  A. Yousef and Z. Jovanovic, "Flow-Based Anomaly Intrusion Detection System using Neural Network," *Computer Engineering Department,University of Belgrae,Belgrade, Serbia*, 2012
[15]  J. Han and M. Kamber, "Data Mining: Concepts and Techniques", *San Diego Academic Press*, 2001.
[16]  L. Alhazzaa and H Mathkour "Intrusion Detection Systems Using Genetic Algorithms," *12th Annual Canadian Information Technology Security Symposium*, 2007.
[17]  C. Kruegel and T. Toth, "*Using Decision Trees to Improve Signature-Based Intrusion Detection*," 43rd annual Southeast regional conference, 2005.
[18]  B. Dong and X. Wang, "*Comparison deep learning method to traditional methods using for network intrusion detection*," in Proc. 8th IEEE Int. Conf. Commun. Softw. Netw., Beijing, China, pp. 581–585, Jun 2016.
[19]  R. Zhao, *et al*., "Deep learning and its applications to machine health monitoring: A survey," *Submitted to IEEE Trans. Neural Netw. Learn. Syst*., 2016.
[20]  S. Hou, *et al*., "*Deep4MalDroid: A Deep learning framework for android malware detection based on linux kernel system call graphs*," *in* Proc. IEEE/WIC/ACM Int. Conf. Web Intell. Workshops, Omaha, NE, USA, pp. 104–111, Oct 2016.
[21]  J. Majidpour, *et al*., "*Interactive tool to improve the automatic image annotation using MPEG-7 and multi-class SVM,*" 7th International Conference on Information and Knowledge Technology, pp. 1-7, 2015.
[22]  J. Majidpour and S. K. Jameel, "*Automatic image annotation base on Naïve Bayes and Decision Tree classifiers using MPEG-7*," 5th Conference on Knowledge-Based Engineering and Innovation, Iran University of Science and Technology, Tehran, Iran, pp.7-12, 2019.
[23]  M. Amanlou and S. M. Mostafavi, "In sillico screening to aim computational efficient inhibitors of caspase-9 by ligand-based pharmacophore modeling," *Medbiotech Journal*, vol. 01, no. 1, pp. 34-41, 2017.
[24]  S. M. Mostafavi1, H. Malekzadeh and M. S. Taskhiri, "In Silico Prediction of Gas Chromatographic Retention Time of Some Organic Compounds on the Modified Carbon Nanotube Capillary Column," *Journal of Computational and Theoretical Nanoscience*, vol. 16, no. 1, pp. 151-156, 2019.
[25]  S. M. Mostafavi, K. Bagherzadeh, M. Amanlou, "A new attempt to introduce efficient inhibitors for Caspas-9 according to structure-based Pharmacophore Screening strategy and Molecular Dynamics Simulations," *Medbiotech Journal*, vol. 1, no. 1, pp. 1-8, 2017.
[26]  A. Ahmadipour, P. Shaibani, S. A. Mostafavi "Assessment of empirical methods for estimating potential evapotranspiration in Zabol Synoptic Station by REF-ET model," *Medbiotech Journal*, vol. 03, no. 1, pp. 1-4, 2019.
[27]  R. Naderi, "Solving a nonlinear Singular Cauchy Problem of Euler- Poisson-Darboux Equation through Homotopy Perturbation Method," *Medbiotech Journal*, vol. 3, no. 1, pp. 29-34, 2019.
[28]  K. Malathi *et al*., "Preterm birth prognostic prediction using Cross domain data fusion," *International Journal of Communication and Computer Technologies*, vol. 7, pp. 10-13, 2019.
[29]  N. H. kumar, *et al*., "A review on Adverse drug reactions monitoring and reporting," *International Journal of Pharmacy Research & Technology*, vol. 9, no. 2, pp. 12-15, 2019.
[30]  E. Mazurova, "Exploratory Analysis of the Factors Affecting Consumer Choice in E-Commerce: Conjoint Analysis," *Journal of Information Systems Engineering & Management*, vol. 2, pp. 12-21, 2017.

**BIOGRAPHIES OF AUTHORS**

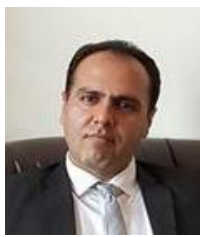

**Jafar Majidpour** was born in Mahabad, West Azerbaijan, Iran in 1982. He received his bachelor degree in Computer Software from Omid Nahavand Institute of Higher Educaation, Nahavand, Hamadan, Iran in 2009. He received Master's degrees in Software Engineering-Software from East Azarbaijan Science Research Branch, Tabriz, Iran in 2013. He has joined University of Raparin from Iraq as an Assistant lecturer in the Basic Education Department since 2013 and he has got lecturer degree in 2019. His primary research interests are machine learning, deep learning, image processing.

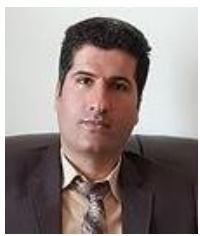

**Hiwa Hassanzadeh** was born in Mahabad, West Azerbaijan, Iran in 1983. He received his bachelor degree in Computer Sciences from Payam-E-Noor university (PNU), Mahabad, Iran in 2009. He received Master's degrees in Software Engineering from (PNU), Tehran, Iran in 2012. He has joined University of Raparin from Iraq as an Assistant lecturer in the Basic Education Department since 2013 and he has got lecturer degree in 2019. His primary research interests are machine learning, deep learning, signal processing, and improve algorithm.